\documentclass[12pt,preprint]{aastex}

\shorttitle{{\it Chandra} \& {\it Swift} observations of HLX-1}
\shortauthors{Webb et al.}

\begin{document}


\title{{\it Chandra} and {\it Swift} follow-up observations of the intermediate mass black hole in ESO243-49}


\author{N. A. Webb, D. Barret, O. Godet}
\affil{Universit\'e de Toulouse, UPS, CESR, 9 Avenue du Colonel Roche, F-31028 Toulouse Cedex 9, France}
\email{natalie.webb@cesr.fr}
\author{M. Servillat}
\affil{Harvard-Smithsonian Center for Astrophysics, 60 Garden Street, MS-67, Cambridge, MA 02138, USA}
\author{S. A.  Farrell}
\affil{Department of Physics and Astronomy, University of Leicester, University
Road, Leicester, LE1 7RH, UK}
\author{S. R.  Oates}
\affil{Mullard Space Science Laboratory/UCL, Holmbury St. Mary, Dorking, Surrey, RH5 6NT, UK}

\begin{abstract}
The brightest Ultra-Luminous X-ray source HLX-1 in the galaxy ESO 243-49 provides strong evidence for the existence of intermediate mass black holes. As the luminosity  and thus the mass estimate depend on the association of HLX-1 with ESO 243-49, it is essential to confirm its affiliation.  This requires follow-up investigations at wavelengths other than X-rays, which in-turn needs an improved source position. To further reinforce the intermediate mass black hole identification, it is necessary to determine HLX-1's environment to establish whether it could potentially form and nourish a black hole at the luminosities observed. Using the High Resolution Camera onboard {\it Chandra}, we determine a source position of RA=01$^h$10$^m$28$\fs$3 and Dec=-46$^\circ$04'22$\farcs$3. A conservative 95\% error of 0.3" was found following a boresight correction by cross-matching the positions of 3 X-ray sources in the field with the 2MASS catalog. Combining all {\it Swift} UV/Optical Telescope {\it uvw2} images, we failed to detect a UV source at the {\it Chandra} position down to a 3 $\sigma$ limiting magnitude of 20.25 mag. However, there is evidence that the UV emission is elongated in the direction of HLX-1. This is supported by archival data from {\it GALEX} and suggests that the far-UV emission is stronger than the near-UV. This could imply that HLX-1  may be situated near the edge of a star forming region.  Using the latest X-ray observations we deduce the mass accretion rate of a 500 M$_{\odot}$ black hole with the observed luminosity and show that this is compatible with such an environment.

\end{abstract}


\section{Introduction}

The brightest Hyper Luminous X-ray source  (HLX-1) was discovered
serendipitously with {\it XMM-Newton} on 2004 November 23 \citep{Farrell09} in the
outskirts of the edge-on spiral galaxy ESO 243--49, at a redshift of
0.0224 \citep{Afonso05}. Its 0.2 -- 10 keV unabsorbed X-ray
luminosity, assuming isotropic emission and the galaxy distance (95
Mpc), exceeded $1.1 \times 10^{42}$ ergs s$^{-1}$. Follow-up
observations with {\it XMM-Newton} and  the {\it Swift} X-ray
telescope (XRT) revealed that the source was variable in X-rays by
more than one order of magnitude, and that luminosity changes were
accompanied by changes in the spectral shape, in a way similar to
Galactic black hole systems \citep{godetapjl09}. As argued before,
either super-Eddington accretion or beaming of the X-ray emission
could account for X-ray luminosities up to $\sim 10^{40}$ergs s$^{-1}$
for stellar mass black holes (10-50 M$_\odot$), but would require
extreme tuning to explain an X-ray luminosity of $10^{42}$ ergs
s$^{-1}$. Hence, HLX-1 is an excellent intermediate mass black hole
(IMBH) candidate \citep{Farrell09}. 

To verify the IMBH nature it is
essential that its association with the host galaxy is confirmed.  To
do this, an improved position is crucial in order to carry out
multi-wavelength follow-up observations.  Subsequently, it is
necessary to determine whether the source resides in an environment
which can provide sufficient material to power such a massive black
hole \citep{milos09}.  Again multi-wavelength observations are the key to
determining whether this object is in a star cluster, a star forming
region or a globular cluster.  As ULXs emit mostly in the X-ray domain, X-ray observations are thus the
key to understanding how much material is required to feed the black
hole as the luminosity is directly related to the mass accretion
rate.  Therefore constant monitoring of the X-ray emission will
allow us to place constraints on the quantity of material necessary in
the neighbouring environment.


Here we present {\it Chandra} observations of HLX-1 with the High
Resolution Camera (HRC) accorded under the Directors Discretionary
Time (DDT) program, which has allowed us to revise and improve the
position of HLX-1, along with {\em Swift} XRT data revealing that the X-ray
luminosity of this source remains above $10^{42}$ ergs
s$^{-1}$, following its recent rebrightening in August 2009 \citep{godetapjl09}.  We discuss the mass accretion rate of HLX-1 using these
observations. We also reveal evidence for the nature of the
environment around HLX-1 through ultra-violet imaging with the {\it
  Swift} UV/Optical Telescope (UVOT); a finding independently
supported by archival {\em Galaxy Evolution Explorer} ($\it{GALEX}$) observations.

\section{Data analysis}
\subsection{The first HRC-I observation}

We obtained a 1 ks DDT observation of HLX-1 with the HRC-I camera onboard
{\it Chandra} on 2009 July 4 (ObsID: 10919).  We extracted all the events in the
detector (i.e. the background) and plotted the number of counts per
energy channel (PI) with the {\scriptsize CIAO} v4.1.1
task {\scriptsize WAVDETECT}. Between PI = 25 and 120 the background is lower and the
sensitivity of the instrument is higher \citep[see
  e.g.][]{cameron}. We thus filtered the event list to include only these energies and performed
source detection with {\scriptsize WAVDETECT} utilising an exposure
map. No source is detected
within the {\it XMM-Newton} error circle of HLX-1, and only 2 sources
are detected in the field of view. The two sources that are detected correspond to known {\it
  XMM-Newton} sources: 2XMM J011050.4-460013  which was similar in
flux to the {\it XMM-Newton} detection of HLX-1, and 2XMM
J010953.9-455538. These sources have respectively 17 and 6 counts in the
HRC-I image, which gives 0.018 and 0.006 ct s$^{-1}$. Therefore,
HLX-1 must have dropped in flux down to at most 0.006 ct s$^{-1}$
(the faintest source detected), compared to 0.03 ct s$^{-1}$ expected
from the flux in the most recent {\it XMM-Newton} observation.  This faint flux is confirmed using {\em Swift} observations taken one month later \citep{godetapjl09}.

\subsection{The second HRC-I observation}

Following the re-brightening of HLX-1 \citep{godetapjl09} we obtained
a second deeper DDT observation of 10 ks with the {\it Chandra} HRC-I
2009 on August 17 (ObsID: 11803).  We generated several images of the
HRC-I field of view with a binning of 4, 8, 16 and 32 pixels, and
performed source detection using {\scriptsize WAVDETECT}.  A total of
11 sources were detected, including a source consistent with the
position of HLX-1 with a net count rate of 0.098 $\pm$ 0.003  ct
s$^{-1}$. We cross-matched the entire source list with the 2MASS
catalogue \citep{skrutskie} and found 3 matching objects which appear
to be the real counterparts to the X-ray sources. A boresight
correction was applied to the X-ray image which lead to a small shift
of 0.285\arcsec. After applying the astrometry correction to the image
{\scriptsize WAVDETECT} was run again. The final corrected position
for HLX-1 was found to be RA = 01$^h$10$^m$28$\fs$3 and Dec = -46$^\circ$04'22$\farcs$3, which
is consistent with the previous derived positions. The error on this
position was then estimated from the 1 $\sigma$ error of the 2MASS
catalogue at that position \citep[0.1\arcsec\ see][]{skrutskie}, the 1
$\sigma$ {\scriptsize WAVDETECT} 2D error (0.030\arcsec), and the root
mean square (RMS) of the alignment of the X-ray image on the 2MASS
reference stars (0.003\arcsec). Combining these errors, a total 95\%
error of 0.3\arcsec\ was derived.

\subsection{The {\it Swift} UVOT data}
The {\it Swift} UVOT observed the field of ESO 243-49 on 2009 August 5, 6, 16,
18, 19 and 20 for a total exposure of 38~ks.
Observations were performed in the {\it uvw2 } ($\sim$1600-2500\AA) filter only.  Data reduction and analysis of the data were done 
using the software release {\sc headas} 6.6.2 and version 20090630 of
the UVOT calibration files.
 At the location
of the core of ESO 243-49 there is an extended object (Figure \ref{fig2}), with some hints of an elongated emission towards the position of HLX-1. No other source is observed above the flux level of this
galaxy at the {\it Chandra} position of HLX-1. 

We cannot determine an upper limit from subtracting the galaxy
contribution at that position as we do not know when or even if HLX-1
stops emitting in the {\it uvw2 } wavelength range. A source at the
X-ray position would only be discernible from the galaxy if the source
is significantly brighter than the emission at that
position. Therefore, to determine an upper limit to the detection of
the ULX above the flux of the galaxy, we first co-added the individual
observations. We then placed 7 circular apertures of 3\arcsec\ radius
on the image, one at the location of the X-ray source and 6 others
around an ellipse on the galaxy of similar galaxy isophotal brightness
to the region containing the source.  Then using a background region
of 20\arcsec\, located in a source free region close to the galaxy, we
determined the magnitude of the galaxy in each aperture using the {\it
  Swift} tool {\sc uvotsource}.  To be compatible with the
calibration, which is determined for a 5\arcsec\ aperture
\citep{poole}, we applied an aperture correction to the count rate
using a table of aperture correction factors contained within the
calibration files. We determined the standard deviation of the
magnitudes determined from the 7 apertures to be 0.10 mag. We then
took the 3 $\sigma$ upper-limit for detecting a source at the {\it
  Chandra} position to be the magnitude determined from the source
region located at the {\it Chandra} position minus 3 times the
standard deviation, which gives a 3 $\sigma$ upper-limit of $20.25$
mag.

\subsection{The {\it GALEX} data}
Given the possible detection of extended emission in the {\it uvw2}, we
retrieved archived data taken by {\it
  GALEX} in the near- (NUV, $\sim$ 1800-2800\AA) and far-UV (FUV,
$\sim$1500-2000\AA) bands. The field of HLX-1 was observed by {\it
  GALEX} as part of the deep survey on 2004 September 27 for $\sim$13
ks in the NUV and $\sim$8 ks in the FUV. A clear extension towards the
location of HLX-1 from the bulge of ESO 243-49 can be seen in both
images (see Figure \ref{fig3}), with the dominant emission occurring
in the FUV. No point source was detected coincident with the position
of HLX-1 in either band. In order to determine magnitude upper limits,
we measured the counts within regions centered on the {\it Chandra}
position with radii corresponding to the 80$\%$ encircled energy radii
(3'' and 4'' for the NUV and FUV respectively). The total count rate
inside six similar sized regions located around the galaxy, at the
same isophotal brightness as the location of HLX-1 were also
determined.  Using a source-free area of the image near the galaxy
with radii of 6'' and 8'' for the NUV and FUV respectively we measured
the total background rates.  These were scaled to the same extraction
region size as used for the source regions.  All the count rates
(both source and background) were then corrected for the
wings of the PSF outside the extraction region. Net count rates were
calculated by subtracting the background rates from the source
rates.  Magnitudes were then calculated for the net rates using the
zero points given in \citet{Morrissey07}. The standard deviation
for the magnitudes from each of the 7 regions were 0.3 and 0.2 mag for
the NUV and FUV respectively.  We then determined the 3 $\sigma$ upper
limits as the magnitudes derived for the region centered on the HLX-1
{\it Chandra} position minus 3 times the standard deviation, in the same way
as for the UVOT data. This gives upper limits of 20.4 mag and 21.4 mag
for the NUV and FUV respectively.

\subsection{The XRT data}

\begin{figure}[!t]
\includegraphics[angle=-90,width=\columnwidth]{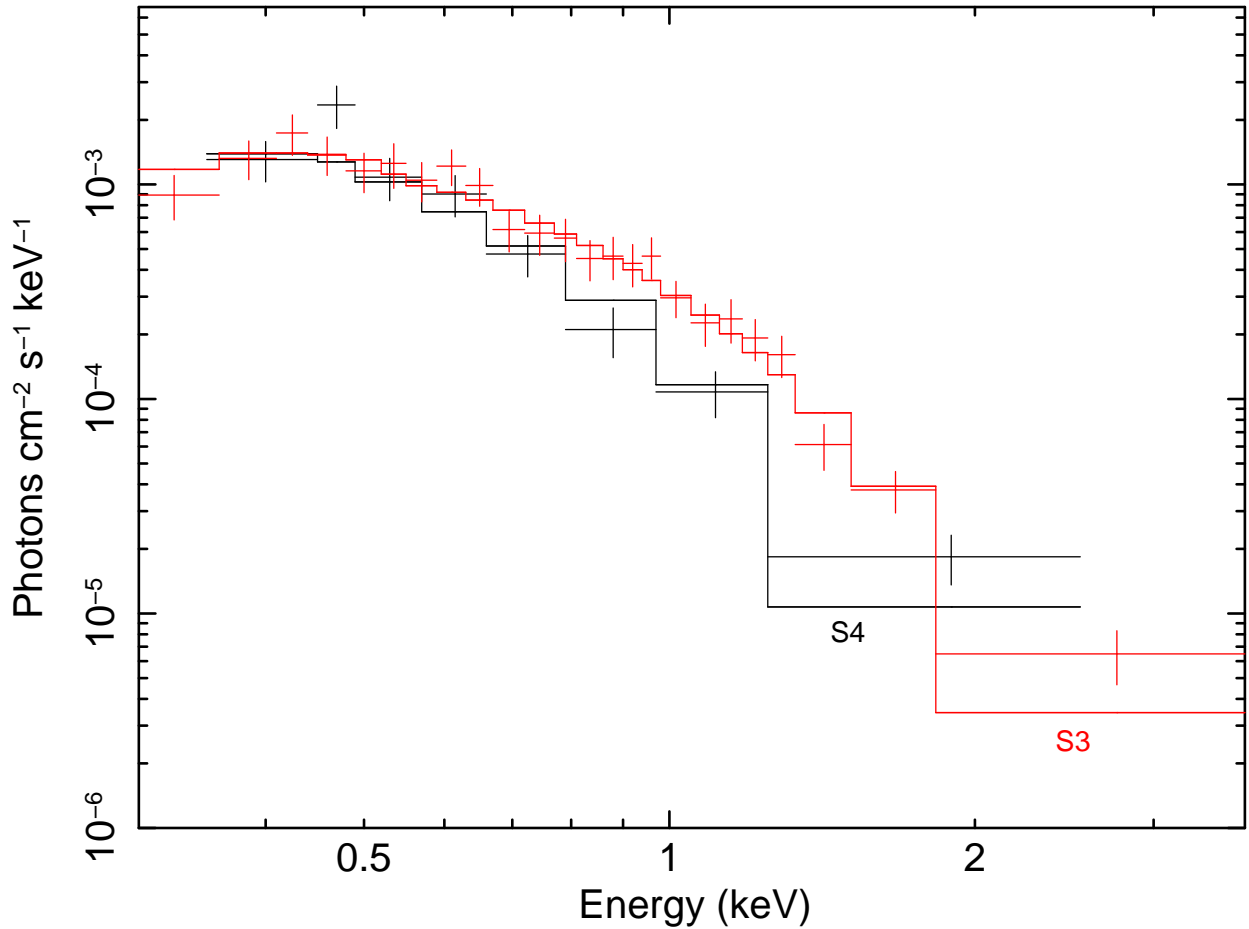}
\caption{{\it Swift}-XRT PC grade 0--12 unfolded spectra of HLX-1: S4
  $=$ 2009-10-02 (black) and S3 $=$ 2009-08-16 (red).  The solid lines
  correspond to the best-fit models. For S4, we used a unique absorbed
  disk black body.  The $n_H$-value was fixed at $4\times 10^{20}$
  cm$^{-2}$ in both cases. }
  \label{fig1}
\end{figure}

\begin{figure}[!h]
\includegraphics[width=\columnwidth]{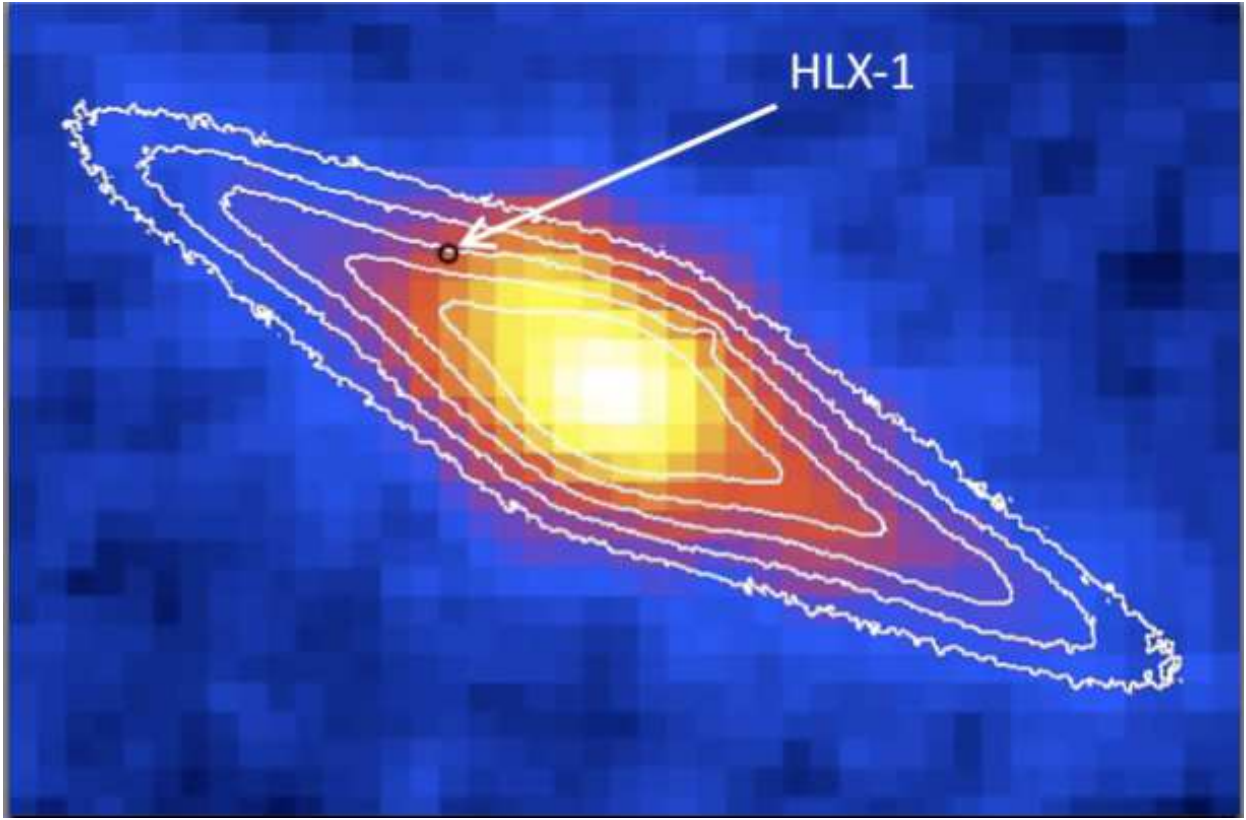}
\caption{{\it Swift}-UVOT un-smoothed {\it uvw2 } image of ESO
  243-49. The white contours show the orientation of the galaxy in the
  J-band (Webb et al. in preparation). The black circle indicated by
  the white arrow is centered on the {\it Chandra} position of HLX-1,
  with the radius representing the 95$\%$ error bounds.}
  \label{fig2}
\end{figure}

\begin{figure*}[!t]
\centerline{\includegraphics[width=\columnwidth]{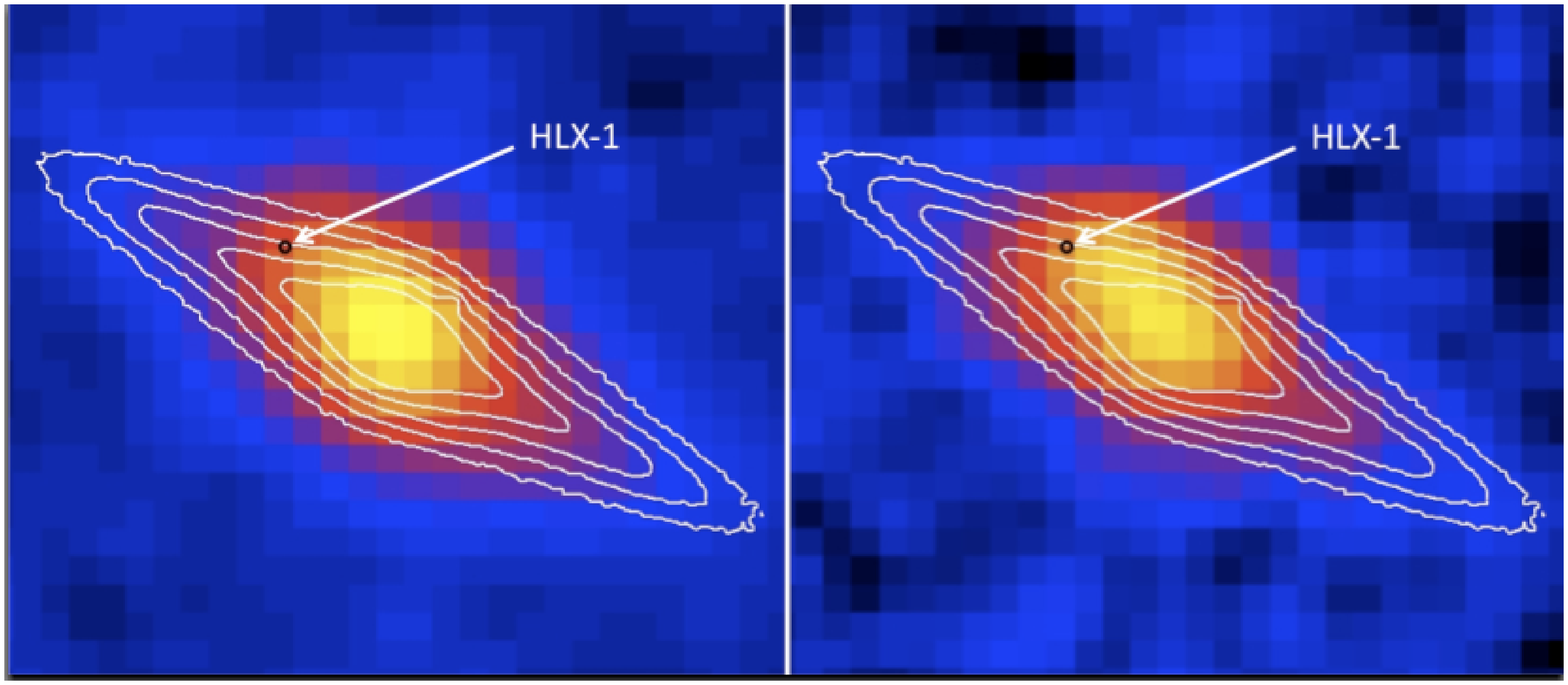}}
\caption{Archival {\it GALEX} near-UV (left) and far-UV (right) images
  of ESO 243-49. Both images have been smoothed using a 2 pixel radius
  Gaussian smoothing function. The white contours show the orientation
  of the galaxy in the J-band (Webb et al. in preparation). The black
  circles indicated by the white arrows are the {\it Chandra} position
  of HLX-1, with the radii indicating the 95\% positional error of
  0.3".  }
  \label{fig3}
\end{figure*}

The third Target-of-Opportunity (ToO) observation of our {\it
  Swift}-XRT monitoring campaign was performed on 2009 October 02
\citep[hereafter designated S4 as it is the fourth of our {\em Swift}
observations of this source, see][]{godetapjl09} for 9.4 ks. All
the {\it Swift}-XRT Photon Counting data were processed using the tool
{\scriptsize XRTPIPELINE} v0.12.3. To analyse the data, we used the
same method as described in \citet{godetapjl09}. This new observation
revealed that the HLX-1 0.3--10 keV count rate dropped from $\sim
3.3\times 10^{-2}$ to $\sim 1.9\times 10^{-2}$ count s$^{-1}$ when
compared to the previous observation taken on 2009 August 16
\citep[S3; see ][]{godetapjl09}. Along with this reduction in count
rate, we also observe a spectral softening as illustrated in
Fig.~\ref{fig1}.  The hardness, defined as the ratio of the 0.3--1 keV
count rate over the 1--10 keV count rate, is $0.26\pm 0.05$ in S4,
compared to a value of $0.42\pm 0.04$ in S3. The errors are quoted at
the 1 $\sigma$ level.  Both a unique absorbed disk black-body (DBB)
and an absorbed power-law (PL) model give comparable fits ($\chi^2/dof
= 9.8/6$ and $\chi^2/dof = 10/6$, respectively). The $n_H$-value was
fixed at $4\times 10^{20}$ cm$^{-2}$ \citep{Farrell09} as the poor
quality of the spectra meant that we were not able to derive a
meaningful value for this quantity, although the allowed values always
included the fixed one. For the DBB model, we obtained a temperature
of $kT = 0.20^{+0.03}_{-0.02}$ keV, consistent with the disc blackbody
temperature found in previous observations with {\em XMM-Newton}
\citep[see][]{Farrell09} and a 0.2--10 keV unabsorbed luminosity of
$1.0\pm 0.3\times 10^{42}$ erg cm$^{-2}$ s$^{-1}$, while for the PL
model we obtained a value for the photon index of $\Gamma = 3.5\pm
0.3$ and a 0.2--10 keV unabsorbed luminosity of
$1.9^{+0.5}_{-0.4}\times 10^{42}$ erg cm$^{-2}$ s$^{-1}$.  A full
discussion of the spectral nature of HLX-1 can be found in \citet{godetapjl09}.   We add here that the spectral state observed for S4
  is intermediate between S1 and S3, which appears to portray smooth
  spectral evolution that resembles that of black hole binaries, supporting
  the interpretation in \citet{godetapjl09}.  However, here we exploit
  only the luminosity so as to constrain mass accretion limits, as
discussed below.


\section{Discussion and conclusions}

Using {\em Chandra} observations of HLX-1, we have determined an
improved position with a 95\% confidence error radius of 0.3\arcsec.
This is sufficiently small that we should eventually be able to
identify the source at other wavelengths, and eventually perform broad
band spectroscopic observations.

Although no point-like source is detected with UV observations at this
position, HLX-1 appears to be situated near the edge of a region of UV
excess stretching from the nucleus of ESO 243-49 towards the {\em Chandra}
position. However, it is
  not yet certain that this emission is related to HLX-1, but the fact
that no similar extended emission is seen in the radio, infra-red,
optical, or X-ray domains \citep[][Webb et al. in preparation]{Farrell09},
makes it unlikely that this is either a foreground or background
source.  Follow-up observations with a higher resolution instrument
would also help to confirm the association and the extended nature, as
the low resolution of {\em GALEX} (5\arcsec\ Full Width Half Maximum
(FWHM)) and the {\em UVOT} (better at 2.9\arcsec\ FWHM) would mean
that it would be difficult to resolve a source at the centre of the
galaxy and a second towards the position of HLX-1.  However, the fact
that this emission appears to be stronger in the FUV could hint
towards star formation taking place in that region, as the UV flux
primarily originates from the photospheres of O- through later-type
B-stars (M$>$3M$_\odot$), and thus measures star formation averaged
over a $\sim$10$^8$ yr timescale \citep[e.g.][]{lee09}.  Starburst
environments are thought to be able to generate IMBHs through runaway
collisions and mergers of massive stars \citep{freitag06}.  Further, the massive stars present in such environments can
supply the necessary material to be accreted onto the black hole to
provide the luminosities observed if one is captured by the black hole
and then proceeds to main sequence Roche-lobe overflow \citep{hopman04}.  An alternative situation is
described by \citet{sun10} where a trail of star formation could
be created by ram-pressure stripping of gas and stars from a dwarf
galaxy which has recently interacted with ESO 243-49. In this case
HLX-1 could have been an intermediate mass black hole which was once
at the centre of the dwarf galaxy, but has now had most of the gas and
stars stripped from it via the gravitational interaction with ESO
243-49.  This may resemble a globular cluster if detected in the
optical domain.  

To determine whether accretion from the medium around the stripped
galaxy is possible, we use the latest X-ray observation of HLX-1,
which has a 0.2--10 keV unabsorbed luminosity of
$1.9^{+0.5}_{-0.4}\times 10^{42}$ erg cm$^{-2}$ s$^{-1}$ if we fit
with a power law model.  This is the highest luminosity that we have
observed over the last five years whilst this source has been bright
(previous observations with {\em Rosat} in the early nineties gave
non-detections, confirming that the source was more than a factor 10
fainter than  in these observations, Webb et al. in preparation). HLX-1 is
not bright in the radio, infra-red, optical or NUV domains \citep[][Webb et al. in preparation]{Farrell09}, so we can assume that the majority of
the emission is in the X-rays and we can use this luminosity (L) to
deduce the approximate mass accretion rate (\.M), where L = $\eta$ \.M
c$^2$ and $\eta$ = GM/Rc$^2$ (M is the mass of the black hole and R
the innermost stable circular orbit i.e. 6 times the Schwarzschild
radius, therefore supposing a non-rotating black hole).  We assume a
mass of 500 M$_\odot$.  We find \.M $\sim$ 2.1 $\times$ 10$^{-4}$
M$_\odot$ yr$^{-1}$.  This is well within the ranges of the matter
available for accretion predicted by \citet{sun10}, therefore
supporting the idea that HLX-1 {\em could} be embedded in a star
forming region and that it may have originated from a stripped dwarf
galaxy.

\acknowledgments

 We thank Neil Gehrels and the {\em Swift team} for according us the
 {\em Swift} observations as well as Harvey Tananbaum and the {\em
   Chandra} team for approving the two {\em Chandra} DDT
 observations. S.A.F. and S.O. acknowledge STFC funding. This research
 has made use of data obtained from the {\it Chandra} Data Archive and
 software provided by the {\it Chandra} X-ray Center. We thank the
 {\it GALEX} collaboration and the Space Telescope Science Institute
 for providing access to the UV images used in this work. {\it GALEX}
 is a NASA Small Explorer Class mission. We thank Simon Rosen for
 valuable discussions and we are grateful to the referee for comments
 which improved this paper.

{\it Facilities:} \facility{{\it Chandra}}, \facility{{\it Swift}}, \facility{{\it GALEX}}


\begin{thebibliography}{}

\bibitem[Afonso et al.(2005)]{Afonso05} Afonso, J. et al. 2005 ApJ, 624,
135

\bibitem[Cameron et al.(2007)]{cameron} Cameron, P.~B., 
Rutledge, R.~E., Camilo, F., Bildsten, L., Ransom, S.~M., 
\& Kulkarni, S.~R.\ 2007, \apj, 660, 587 


\bibitem[Farrell et al.(2009)]{Farrell09} Farrell, S. A., Webb, N. A.,
    Barret, D., Godet, O. \& Rodrigues, J. M.  2009, Nature, 460, 73


\bibitem[Freitag, Rasio \& Baumgardt(2006)]{freitag06} Freitag, M., 
Rasio, F.~A., \& Baumgardt, H., 2006, \mnras, 368, 121 

\bibitem[Godet et al.(2009)]{godetapjl09} Godet, O., Barret, D., 
Webb, N.~A., Farrell, S.~A., \& Gehrels, N.\ 2009, ApJL, 705, 109

\bibitem[Hopman, Portegies Zwart \& Alexander(2004)]{hopman04} Hopman, C., Portegies Zwart, S. F., \& Alexander, T., 2004, ApJ, 604, 101

\bibitem[Lee et al.(2009)]{lee09} Lee, J.C., de Paz, A. G., Temonti, C., \& et al., 2009, ApJ, 706, 599

\bibitem[Miller  \& Colbert(2004)]{miller04} Miller, M.~C., \& Colbert, E.~J.~M.\ 2004, International Journal of Modern Physics D, 13, 1 


\bibitem[Milosavljevi{\'c} et al.(2009)]{milos09} Milosavljevi{\'c}, M., Couch, S.~M., \& Bromm, V.\ 2009, ApJL, 696, L146 

\bibitem[Morrissey et al.(2007)]{Morrissey07} Morrissey, P., et 
al.\ 2007, ApJSS, 173, 682 

\bibitem[Poole et al.(2008)]{poole} Poole, T.~S., et al.\ 
2008, \mnras, 383, 627 

\bibitem[Skrutskie et al.(2006)]{skrutskie} Skrutskie, M.~F., et 
al.\ 2006, \aj, 131, 1163 


 \bibitem[Sun et al.(2010)]{sun10} Sun, M., Donahue, M., Roediger, E., \& et al. 2010, ApJ, 708, 946


\end{thebibliography}
\end{document}